\documentclass{article}
\usepackage{graphicx,amsmath,amsfonts,amssymb,amsbsy,amsgen}
\def\bm#1{\mathchoice
 {\mbox{\boldmath$\displaystyle#1$}}%
 {\mbox{\boldmath$#1$}}%
 {\mbox{\boldmath$\scriptstyle#1$}}%
 {\mbox{\boldmath$\scriptscriptstyle#1$}}}
\newcommand{\ud}{\mathrm{d}}
\newcommand{\ue}{\mathrm{e}}

\newcommand{\als}{\alpha^*}
\newcommand{\alt}{\tilde{\alpha}}
\newcommand{\altsu}{\xi_{\textit{\scriptsize I}}}
\newcommand{\altsd}{\xi_{\textit{\scriptsize II}}}
\newcommand{\altst}{\xi_{\textit{\scriptsize III}}}

\author{Marco Martins Afonso\\
 \small Department of Physics of Complex Systems,\\[-0.1cm]\small Weizmann Institute of Science, Rehovot (Israel)}
\title{Fluid-particle separation in the Batchelor regime with telegraph model of noise}
\date{\today}

\begin{document}

\maketitle

\begin{abstract}
We study the statistics of the relative separation between two
fluid particles in a random flow. We confine ourselves to the
Batchelor regime, i.e.~we only examine the evolution of distances
smaller than the smallest active scale of the flow, where the
latter is spatially smooth. The Lagrangian strain is assumed as
given in its statistics and is modelled by a telegraph noise. This
is a stationary random Markov process, which can only take two
values with known transition probabilities. The presence of two
independent parameters (intensity of velocity gradient and flow
correlation time) allows the definition of their ratio as the Kubo
number, whose infinitesimal and infinite limits describe the
delta-correlated and quasi-deterministic cases, respectively.
However, the simplicity of the model enables us to write closed
equations for the interparticle distance in the presence of a
finite-correlated, i.e.~coloured, noise.\\
In 1D, the flow is locally compressible in every single
realization, but the average `volume' must keep finite. This
provides us with a mathematical constraint, which physically
reflects in the fact that, in the Lagrangian frame, particles
spend longer time in contracting regions than in expanding ones.
Imposing this condition consistently, we are able to find
analytically the long-time growth rate of the
interparticle-distance moments and, consequently, the senior
Lyapunov exponent, which coherently turns out to be negative.
Analysing the large-deviation form of the joint probability
distribution, we also show the exact expression of the Cram\'er
function, which happens to satisfy the well-known fluctuation
relation despite the time irreversibility of the
strain statistics.\\
The 2D incompressible case is also studied. After showing a simple
generalization of the 1D situation, we concentrate ourselves on
the general isotropic case: the evolution of the linear and
quadratic components is analysed thoroughly, while for higher
moments, due to high computational cost, we focus on a restricted,
though exact, dynamics. As a result, we obtain the moment
asymptotic growth rates and the Lyapunov exponent (positive) in
the two above-mentioned limits, together with the leading
corrections. The quasi-deterministic limit turns out to be
singular, while a perfect agreement is found with the
already-known delta-correlated case.
\end{abstract}

\section{Introduction}

Understanding the behaviour of the relative separation between two
fluid particles in a generic flow is a difficult task of paramount
importance. The most intuitive application of this concept is
represented by the passive scalar problem, where the Eulerian
description in terms of fields has its Lagrangian counterpart just
in the study of fluid trajectories \cite{FGV01}.\\
While in the one-dimensional situation general considerations and
results can be carried out (at least for smooth flows), only few
cases are known to be solvable in higher dimension
\cite{BF99,MK99}; namely, a short-correlated strain (fully
solvable), a 2D slow strain \cite{CFKL95} and the
large-dimensionality case \cite{FKL98}.

In this paper we describe another situation which allows exact
analytical calculations: the telegraph-noise model for the
velocity gradient in the Batchelor regime \cite{B59}. This means
that the relative separation between two fluid particles,
$\bm{R}(t)$, evolves according to
\begin{equation} \label{sigma}
 \ud_t\bm{R}(t)=\sigma(t)\bm{R}(t)\;,
\end{equation}
and the Lagrangian strain (scalar or matrix) $\sigma(t)$ is
assumed as properly made up of a telegraph noise \cite{SL78}. We
shall study both the 1D (compressible) and the incompressible 2D
cases.

The telegraph noise is a stationary random process, $\alpha(t)$,
which satisfies the Markov property and only takes two values,
$a_1$ and $a_2$. The probability, per unit time, of passing from
the latter state to the former (or viceversa) is given by $\nu_1$
($\nu_2$, respectively). In what follows, we shall consider some
special cases, for which simple formulas hold.\\
(i) If $a_1=-a_2$, then the process itself is stochastic but its
square is deterministic, keeping the constant value
$\alpha^2(t)=a^2$ (with $a=|a_1|=|a_2|$). We shall denote such
processes with a star, e.g.~$\als(t)$.\\
(ii) If the average $\langle\alpha\rangle=(\nu_1a_1+\nu_2a_2)/\nu$
vanishes (with $\nu=\nu_1+\nu_2$), then the autocorrelation takes
the form
$\langle\alpha(t)\alpha(t')\rangle=\ue^{-\nu|t-t'|}(\nu_1a_1^2+\nu_2a_2^2)/\nu$.
For any analytical functional $F[\alpha]$ (function of time), a
simple ``formula of differentiation'' \cite{SL78} then holds in
this case:
\begin{equation} \label{fod}
 \ud_t\langle\alpha(t)F[\alpha]\rangle=\langle\alpha(t)\ud_tF[\alpha]\rangle-\nu\langle\alpha(t)F[\alpha]\rangle\;.
\end{equation}
We shall denote such processes with a tilde, e.g.~$\alt(t)$. Note
that it is always possible to reduce to this case from a generic
telegraph noise (in particular from one $\als(t)$), simply by
subtracting from the latter its mean value.\\
(iii) If both the properties in (i) and (ii) hold simultaneously,
the maximum level of simplification is reached. For the sake of
simplicity, we shall denote such processes not with
$\tilde{\alpha}^*(t)$ but rather with $\xi(t)$.

This type of noise has often been used to mimic more general
colored noise in stochastic theories and has also been applied,
e.g., to optics (see \cite{KWW91} and references therein). An
interesting application of the ``telegraph concept'' to turbulence
can be found in \cite{SB06}.

The paper is organized as follows: in section \ref{1d} we will
study the 1D case. In section \ref{2d} we will move to 2D, first
describing a simple peculiar situation (the hyperbolic flow,
\ref{hypf}) and then the general incompressible isotropic case;
the latter will be analysed at first in its complete form, then
(respectively in subsections \ref{erd} and \ref{qdcc}) by means of
a simplified dynamics and eventually in the quasi-delta-correlated
limit. Conclusions follow in section \ref{conc}. The appendices
are devoted to the quasi-deterministic limit (\ref{app1}) and to
some calculation details (\ref{app2}).

\section{1D case} \label{1d}

The simplest meaningful model, making use of the telegraph noise,
for the 1D case, consists in assuming a strain function with the
property (i), i.e.
\begin{equation} \label{sas}
 \sigma(t)=\als(t)\;.
\end{equation}
Indeed, the problem is intimately compressible locally,
i.e.~within every single single realization of the noise, but we
want the average (1D) ``volume'' not to explode or shrink
asymptotically in time. If we denote the long-time evolution of
the moments of the interparticle distance with\footnote{It is easy
to show that the dynamics defined by (\ref{sigma}) is such that
$R(t)$ always keeps the same sign. Therefore, throughout this
section, negative separations can also be considered, but they are
to be dealt with either in absolute value or as the ratio
$R(t)/R(0)$.}
\begin{equation} \label{gamma}
 \langle R^n(t)\rangle\stackrel{t\to+\infty}{\sim}\ue^{\gamma(n)t}\;,
\end{equation}
this means that we have to impose the constraint
\begin{equation} \label{g10}
 \gamma(1)=0\;.
\end{equation}
From (\ref{sigma}) and (\ref{sas}), it is straightforward to
obtain the system of the two first-order ordinary differential
equations which describe the time evolution of moments. It is
indeed sufficient to multiply the starting equation by the noise
itself (taken at the same time) in order to get a closed system,
thanks to the simplicity of the telegraph form. Namely, if we
define $\alpha_0\equiv\langle\als(t)\rangle$ and rewrite
accordingly
\begin{equation} \label{alst0}
 \als(t)=\alt(t)+\alpha_0\;,
\end{equation}
we obtain
\begin{equation} \label{rn0}
 \ud_tR^n(t)=nR^{n-1}(t)\ud_tR(t)=n\als(t)R^n(t)=n[\alt(t)+\alpha_0]R^n(t)
\end{equation}
and thus
\begin{equation} \label{rn}
 \ud_t\langle R^n(t)\rangle=n\langle\alt(t)R^n(t)\rangle+n\alpha_0\langle R^n(t)\rangle\;.
\end{equation}
We must then look for the evolution equation of
$\langle\alt(t)R^n(t)\rangle$, which is readily done by exploiting
(\ref{fod}):
\begin{equation} \label{arn1}
 \ud_t\langle\alt(t)R^n(t)\rangle=\langle\alt(t)\ud_tR^n(t)\rangle-\nu\langle\alt(t)R^n(t)\rangle\;.
\end{equation}
Taking (\ref{rn0}) into account, the first term on the right-hand
side becomes
\begin{eqnarray*}
 \langle\alt(t)\ud_tR^n(t)\rangle\!\!&\!\!=\!\!&\!\!n\langle\alt^2(t)R^n(t)\rangle+n\alpha_0\langle\alt(t)R^n(t)\rangle\\
 &\!\!=\!\!&\!\!n\langle[\als(t)-\alpha_0]^2R^n(t)\rangle+n\alpha_0\langle\alt(t)R^n(t)\rangle\\
 &\!\!=\!\!&\!\!na^2\langle R^n(t)\rangle+n\alpha_0^2\langle R^n(t)\rangle-2\alpha_0\langle[\alt(t)+\alpha_0]R^n(t)\rangle\\
 &&\!\!+n\alpha_0\langle\alt(t)R^n(t)\rangle\\
 &\!\!=\!\!&\!\!na^2\langle R^n(t)\rangle-n\alpha_0^2\langle R^n(t)\rangle-n\alpha_0\langle\alt(t)R^n(t)\rangle\;,
\end{eqnarray*}
where we made use of the property stated in (i). Consequently,
(\ref{arn1}) rewrites as
\begin{equation} \label{arn2}
 \ud_t\langle\alt(t)R^n(t)\rangle=n(a^2-\alpha_0^2)\langle R^n(t)\rangle-(\nu+n\alpha_0)\langle\alt(t)R^n(t)\rangle\;.
\end{equation}
Equations (\ref{rn}) and (\ref{arn2}) constitute the system we
were looking for. It is easy to recast the latter as a
second-order differential equation for the quantity $\langle
R^n(t)\rangle$ alone, which can then be solved exactly:
\begin{eqnarray} \label{errenne}
 &\ud_t^2\langle R^n(t)\rangle+\nu\ud_t\langle R^n(t)\rangle-n(na^2+\nu\alpha_0)\langle R^n(t)\rangle=0\nonumber\\
 &\Longrightarrow\ \langle R^n(t)\rangle=b_n\ue^{-(\sqrt{\nu^2+4n^2a^2+4n\nu\alpha_0}+\nu)t/2}+\beta_n\ue^{(\sqrt{\nu^2+4n^2a^2+4n\nu\alpha_0}-\nu)t/2}\nonumber\\
 &\Longrightarrow\ \gamma(n)=(\sqrt{\nu^2+4n^2a^2+4n\nu\alpha_0}-\nu)/2
\end{eqnarray}
Constraint (\ref{g10}) applied to (\ref{errenne}) implies
\begin{equation} \label{alfagamma}
 \alpha_0=-\frac{a^2}{\nu}\ \Longrightarrow\ \gamma(n)=\frac{\sqrt{\nu^2+4a^2n(n-1)}-\nu}{2}\;.
\end{equation}
The latter is a convex function with asymptotic behaviour
$\gamma(n)\stackrel{n\to\pm\infty}{\sim}\pm an$ and symmetric with
respect to the quadratic minimum
\[\gamma(n)\stackrel{n\sim1/2}{\sim}\frac{\sqrt{\nu^2-a^2}-\nu}{2}+\frac{a^2}{\sqrt{\nu^2-a^2}}\left(n-\frac{1}{2}\right)^2\;.\]
The mean value $\alpha_0$ turns out to be negative but its modulus
cannot clearly exceed $a$, which means that only situations with
$a/\nu$ (the analogue of the Kubo number) smaller than unity are
physically relevant in this picture. Another interesting
consequence is the relation $\nu_1-\nu_2=-a$, which reduces from
three to two the number of free parameters in the original
definition (\ref{sas}) of the strain. This means that the lower
level of the noise persists longer: in the Lagrangian frame,
indeed, particles spend longer time in contracting regions than in
expanding ones, thus enhancing the weight of the former in the
statistics. Accordingly, the Lyapunov exponent, computed as the
derivative of (\ref{alfagamma}) taken in the origin, turns out to
be negative \cite{R99,BFF01}:
\begin{equation} \label{lambda}
 \lambda=\gamma'(0)=-a^2/\nu\;.
\end{equation}

The joint probability density function $P(R,t)$ can also be
investigated. It satisfies the second-order partial differential
equation \cite{SL78}
\begin{equation} \label{prob}
 \partial_t^2P+\nu\partial_tP-a^2\partial_R[R\partial_R(RP)]+a^2\partial_R(RP)=0\;.
\end{equation}
Equation (\ref{prob}) has no stationary solutions, as both
$P=\textrm{const.}$ and $\propto R^{-1}$ are non-normalizable. It
can be solved by means of a saddle-point evaluation after variable
separation, which amounts to look for the large-deviation
asymptotic form
\begin{equation} \label{pro}
 P(R,t)\stackrel{t\to+\infty}{\sim}\ue^{-tH(X)}\;,
\end{equation}
where $X\equiv t^{-1}\ln[R/R(0)]$ can only belong (because of
(\ref{sigma})) to the interval $[-a,a]$. The Cram\'er function
$H(X)$ is simply the Legendre transform of $\gamma(n)$:
\begin{equation} \label{cramer}
 H(X)=\frac{\nu}{2}+\frac{X}{2}-\frac{\sqrt{(\nu^2-a^2)(a^2-X^2)}}{2a}\;.
\end{equation}
It is a convex function vanishing quadratically at the minimum,
represented by the Lyapunov exponent,
$H(X)\stackrel{X\sim\lambda}{\sim}(X-\lambda)^2\nu^3/4a^2(\nu^2-a^2)$,
and it approaches vertically the boundaries of its compact domain:
\[H(X)\stackrel{X\sim\pm a}{\sim}\frac{\nu\pm a}{2}-\sqrt{\frac{\nu^2-a^2}{2a}}\sqrt{a\mp X}\;.\]
Moreover, it satisfies the Evans--Searles--Gallavotti--Cohen
fluctuation relation $H(X)-H(-X)=X$ \cite{ES94,GC95} despite the
time irreversibility of the strain statistics \cite{FF04}.\\
Plugging (\ref{cramer}) into (\ref{pro}), we obtain the final
result\footnote{It is worth mentioning what would happen if one
did not impose the constraint (\ref{g10}) and assumed a strain
$\sigma(t)=\xi(t)$ with the property described in (iii). In this
case, $\gamma(n)=(\sqrt{\nu^2+4n^2a^2}-\nu)/2$ and the Lyapunov
exponent would vanish. The Cram\'er function would be
$H(X)=\nu/2-\nu\sqrt{a^2-X^2}/2a$, so that
\[P(R,t)\stackrel{t\to\infty}{\sim}|R|^{-1}\exp\Big\{-\Big[1-\sqrt{1-\ln^2[R/R(0)]/a^2t^2}\ \Big]\nu t/2\Big\}\]
would be the solution to the equation for the joint probability
distribution, similar to (\ref{prob}) but without the last term.
\label{note}}
\[P(R,t)\stackrel{t\to+\infty}{\sim}\frac{\ue^{-\left[1-\sqrt{(1-a^2/\nu^2)\{1-\ln^2[R/R(0)]/a^2t^2\}}\right]\nu t/2}}{\sqrt{|R|}}\;.\]

\section{2D case} \label{2d}

In a 2D environment it is possible to consistently impose the
incompressibility constraint locally, for every single realization
of the noise. For the sake of simplicity, we shall only consider
this situation, which implies the strain matrix to be traceless.\\
Before studying the general isotropic case, we would like to show
a simpler instance of flow, namely the hyperbolic one, where a
simplification is possible.

\subsection{Hyperbolic flow} \label{hypf}

Let us consider the incompressible irrotational flow described by
the diagonal strain matrix
\[\sigma_{ij}(t)=\left(\begin{array}{cc}\xi(t)&0\\0&-\xi(t)\end{array}\right)\;.\]
The streamlines are represented by fixed equilateral hyperbolae,
which are however travelled by fluid particles in
randomly-reversing direction. For the sake of simplicity the
velocity gradients are assumed to have equal reversal
probabilities and constant absolute values, which justifies the
use of the telegraph signal $\xi(t)$ with the properties
described in (iii).\\
From the analytical point of view, the quantities of our interest
which can be investigated are of the type
\begin{equation} \label{rnk}
 r_{n,k}(t)\equiv R_1^k(t)R_2^{n-k}(t)\;,
\end{equation}
with integer $n\ge k\ge0$. By means of straightforward algebra,
one gets the evolution
\[\langle r_{n,k}(t)\rangle=B_{n,k}\ue^{-\left[\sqrt{\nu^2+4(2k-n)^2a^2}+\nu\right]t/2}+\mathcal{B}_{n,k}\ue^{\left[\sqrt{\nu^2+4(2k-n)^2a^2}-\nu\right]t/2}\;,\]
which implies
\[\langle R^n(t)\rangle\stackrel{t\to+\infty}{\sim}\ue^{(\sqrt{\nu^2+4n^2a^2}-\nu)t/2}\]
for positive even $n$. Therefore, by extrapolation,
\[\gamma(n)=\frac{\sqrt{\nu^2+4n^2a^2}-\nu}{2}\ \Longrightarrow\ \lambda=0\;,\]
which physically reflects the fact that the velocity changes its
sign with equal probabilities in both directions, thus preventing
the interparticle distance from exploding exponentially.\\
The equation for the joint probability distribution $P(\bm{R},t)$
corresponding to (\ref{prob}) reads
\begin{equation}
 \partial_t^2P+\nu\partial_tP-a^2\partial_{R_i}[R^{\dag}_i\partial_{R_j}(R^{\dag}_jP)]=0\;,
\end{equation}
with
$\bm{R}^{\dag}\equiv\left(\begin{array}{c}R_1\\-R_2\end{array}\right)$.
It is evident that, at long times, such probability must factorize
into a product of single-coordinate probabilities, in the sense
that the problem can be reduced to a pair of independent,
one-dimensional problems described in footnote \ref{note}.
Exploiting this fact, we obtain the final result
\[P(\bm{R},t)\stackrel{t\to\infty}{\sim}\ue^{-\left(2-\sqrt{1-\ln^2[R_1/R_1(0)]}/a^2t^2-\sqrt{1-\ln^2[R_2/R_2(0)]}/a^2t^2\right)\nu t/2}/|R_1R_2|\;.\]

\subsection{General isotropic flow}

We now turn to analyse the general isotropic flow described by the
strain matrix
\[\sigma_{ij}(t)=\left(\begin{array}{cc}
 \altsu(t)&\altsd(t)+\sqrt{2}\altst(t)\\
 \altsd(t)-\sqrt{2}\altst(t)&-\altsu(t)
\end{array}\right)\]
The noises $\altsu(t)$, $\altsd(t)$ and $\altst(t)$ are
independent of one another but share the properties in (iii) with
the same coefficients $a$ and $\nu$, which can now range from $0$
to $\infty$ independently. Differently from the 1D case (see
footnote \ref{note}), we will see that this is enough to
mimic a realistic situation. The matrix $\sigma$ thus turns out to
be singular and nilpotent: $\det\sigma=0=\sigma^2$.\\
By means of simple manipulations, analogue to those described for
the 1D case, one can write down a closed system for the
coordinates $r_{n,k}$ defined in (\ref{rnk}) at every $n$, in the
form
\begin{equation} \label{ra}
 \ud_t\langle\mathcal{R}^{(n)}_{\iota}\rangle=\mathtt{A}^{(n)}_{\iota\kappa}\langle\mathcal{R}^{(n)}_{\kappa}\rangle\;.
\end{equation}
The column vector $\mathcal{R}^{(n)}$ has dimension $8(n+1)$,
being made up of: the $n+1$ components of $r_{n,\cdot}$ itself,
the $3(n+1)$ components of $r_{n,\cdot}$ times one noise, the
$3(n+1)$ components of $r_{n,\cdot}$ times two (distinct) noises,
and finally the $n+1$ components of $r_{n,\cdot}$ times all the
three noises. No other quantity is required to close the system,
because as soon as a noise appears as square it becomes a
deterministic quantity and can be taken out of the statistical
average. The matrix $\mathtt{A}^{(n)}$ is thus of order
$8(n+1)\times8(n+1)$ and turns out to be dependent only on $a$ and
$\nu$.\\
In order to take also rotation into account properly, the presence
of three noises is necessary and sufficient; one can easily
understand that such number would sharply increase in higher
dimensions, thus strongly enhancing the number of components in
$\mathcal{R}^{(n)}$: this is why we confine ourselves to the 2D
case.\\
To study the long-time evolution of $\langle r_{n,k}\rangle$, one
should then look for the eigenvalues of $\mathtt{A}^{(n)}$ and
identify the one with largest positive real part. This is not an
easy task from the numerical point of view for growing $n$, and it
does not look feasible analytically for a generic $n$, even if the
matrix itself can be easily written down as a function of $n$ (see
appendix \ref{app2} for details). In particular, we performed the
full calculation and we found the whole spectrum for $n=1$ and
$n=2$.\\
The evolution of the linear components $\langle R_1\rangle$ and
$\langle R_2\rangle$ is very easy to describe: the largest
eigenvalue is $0$, which means that they do not show any
exponential growth at large times; however, as they can change
sign, this does not yield any information on $\langle R\rangle$,
i.e.~on $\gamma(1)$.\\
The situation is much more meaningful for quadratic components
($R_1^2$, $R_1R_2$, $R_2^2$), because in this case the largest
eigenvalue is given by
\begin{equation} \label{mu}
 \mu_2=-\nu+\frac{{\nu}^2}{\sqrt[3]{216a^2\nu+3\sqrt{5184a^4\nu^2-3\nu^6}}}+\frac{\sqrt[3]{216a^2\nu+3\sqrt{5184a^4\nu^2-3\nu^6}}}{3}\;,
\end{equation}
which is always positive and behaves asymptotically as
\begin{equation} \label{as}
 \mu_2=\left\{\begin{array}{ll}
  \displaystyle8\frac{a^2}{\nu}-96\frac{a^4}{\nu^3}\left[1+O\left(\frac{a}{\nu}\right)^2\right]&\displaystyle\textrm{for }\frac{a}{\nu}\to0\\[0.3cm]
  \displaystyle2\sqrt[3]{2a^2\nu}-\nu\left[1+O\left(\frac{\nu}{a}\right)^{2/3}\right]&\displaystyle\textrm{for }\frac{a}{\nu}\to\infty\;.
 \end{array}\right.
\end{equation}
The eigenvalue $\mu_2$ expresses the exponential growth rate of
$\langle R^2\rangle=\langle R_1^2\rangle+\langle R_2^2\rangle$ and
thus coincides with the point $\gamma(2)$ on the curve
$\gamma(n)$.

\subsubsection{Exact reduced dynamics} \label{erd}
As already pointed out, this general process is computationally
very demanding and therefore has not been carried out for $n\ge3$.
The following step thus consists in trying to find a reduced
dynamics, i.e.~a subset of components of $\mathcal{R}^{(n)}$ such
that the corresponding rows in $\mathtt{A}^{(n)}$ are nonzero only
in columns whose index has been taken into account in the reduced
subset.\\
The first hint into this direction is provided by a more careful
look at the matrices $\mathtt{A}^{(n)}$, or equivalently at the
corresponding system of equations. For instance, taking
$\mathtt{A}^{(1)}$ into account, one can see (from appendix
\ref{app2}) that the rows number 1, 3, 6, 8, 10, 12, 13 and 15
form a closed system, and the same is for the remaining 8 rows: we
have thus reduced from a single $16\times16$ problem to a couple
of $8\times8$ independent problems. The same half-splitting can be
easily verified directly for $\mathtt{A}^{(2)}$ and is expected to
take place also for higher $n$, thus contributing to a strong
simplification.\\
However, a much heavier simplification can be obtained, even doing
without any approximation and thus considering exact relations.
Here, the key idea is that we are not interested in studying the
time behaviour in itself of every single component of
$\mathcal{R}^{(n)}_{\iota}$ with $\iota>n+1$ (i.e.~of the
quantities constructed by multiplying the coordinates $r_{n,k}$ by
some noises), but we simply need some combinations of them in
order to describe correctly the dynamics of the first $n+1$
components; moreover, not necessarily all the latter quantities
are of our interest, but only a combination or a subset of them.\\
To be more specific, let us consider again $\mathtt{A}^{(1)}$.
Looking at the coefficients which appear in the first row, which
describes the evolution of $R_1$, it is natural to consider the
following linear combination: third row, plus sixth row, plus
eighth row times $\sqrt{2}$. In other words, we are noticing that
the left product of $\mathtt{A}^{(1)}$ with the row vector
$x_0\equiv(1,0,\ldots,0)$ gives the vector
$x_1\equiv(0,0,1,0,0,1,0,\sqrt{2},0,\ldots,0)$ times $a$. We thus
have to perform $x_1\cdot\mathtt{A}^{(1)}$, and it can easily be
seen that the result is $-\nu x_1$. Basically, we are down to a
$2\times2$ matrix (note that $2=1+1=n+1$) with zeros on one line
and $(a,-\nu)$ on the other: its eigenvalues are trivially 0 and
$-\nu$; 0 being the larger one, in accordance with the result from
the complete form. It is clear that the same result also holds
considering the evolution of $R_2$ (second line of
$\mathtt{A}^{(1)}$), i.e.~$x_0\equiv(0,1,0,\ldots,0)$ and
$x_1\equiv(0,0,0,-1,1,0,-\sqrt{2},0,\ldots,0)$.\\
Let us turn to $\mathtt{A}^{(2)}$. Our aim is to find the
behaviour of $R^2$, therefore let us begin with adding up the
first and third lines, corresponding to $R_1^2$ and $R_2^2$
respectively. In formulas, if $y_0\equiv(1,0,1,0,\ldots,0)$, we
have $y_0\cdot\mathtt{A}^{(2)}=2ay_1$, with
$y_1\equiv(0,0,0,1,0,-1,0,2,0,\ldots,0)$. Then we compute
$y_1\cdot\mathtt{A}^{(2)}=4ay_0-\nu y_1+2\sqrt{2}ay_2$, where
$y_2\equiv(0,\ldots,0,2,0,-1,0,1,0,0,0)$. Finally, we are able to
close the system with
$y_2\cdot\mathtt{A}^{(2)}=-2\sqrt{2}ay_1-2\nu y_2$. The resulting
$3\times3$ matrix (notice that $3=2+1=n+1$) is thus
\[\left(\begin{array}{ccc}
 0&2a&0\\
 4a&-\nu&2\sqrt{2}a\\
 0&-2\sqrt{2}a&-2\nu
\end{array}\right)\]
and its eigenvalues satisfy
\begin{equation} \label{red}
 \mu_2^3+3\nu\mu_2^2+2\nu^2\mu_2-16a^2\nu=0\;,
\end{equation}
which gives the value of $\mu_2$ reported in (\ref{mu}) as the
solution with largest positive real part. It is also worth
noticing that the highest nonzero component in the vectors
$y_0,y_1,y_2$ is the last but three, which means that the final 3
components of $\mathcal{R}^{(2)}$ (which correspond to the
coordinates $r_{2,k}$ times all three noises) are not involved at
all in the dynamics of $\langle R^2\rangle$.\\
Having tested this simplification to $(n+1)\times(n+1)$ on the
already-solved cases $n=1$ and $n=2$, we now turn to higher $n$,
but we confine ourselves to even values. Indeed,
\begin{equation} \label{binom}
 R^n=(R_1^2+R_2^2)^{n/2}=\sum_{k=0,2,\ldots,n}\binom{n/2}{k/2}r_{n,k}
\end{equation}
by definition for positive even $n$, thus the corresponding points
$\gamma(n)$ can be extracted from the knowledge of $n/2+1$ of the
first $n+1$ components in $\mathcal{R}^{(n)}$. On the contrary,
this is not the case for the odd $n$'s, which behave similarly to
non-natural $n$'s because they involve nonlinear operations (in
this case a square root) that do not commute with the statistical
average $\langle\cdot\rangle$.\\
In particular, we analysed the values $n=4$, $n=6$ and $n=8$.
Adopting a technique similar to the one described in the previous
paragraphs, one gets matrices of order 5, 7 and 9 respectively
(i.e.~$n+1$), whose characteristic polynomials are:
\begin{eqnarray*}
 &\mu_4^5+8\nu\mu_4^4+23\nu^2\mu_4^3+(28\nu^3-96a^2\nu)\mu_4^2+(12\nu^4-336a^2\nu^2)\mu_4\\
 &+(-288a^2\nu^3-1152a^4\nu)\;,
\end{eqnarray*}
\begin{eqnarray*}
 &\mu_6^7+11\nu\mu_6^6+49\nu^2\mu_6^5+(113\nu^3-96a^2\nu)\mu_6^4+(142\nu^4-768a^2\nu^2)\mu_6^3\\
 &+(92\nu^5-2208a^2\nu^3-13440a^4\nu)\mu_6^2+(24\nu^6-2688a^2\nu^4-44160a^4\nu^2)\mu_6\\
 &+(-1152a^2\nu^5-34560a^4\nu^3-230400a^6\nu)
\end{eqnarray*}
and
\begin{eqnarray*}
 &\!\!\mu_8^9+12\nu\mu_8^8+60\nu^2\mu_8^7+(128a^2\nu+162\nu^3)\mu_8^6+(255\nu^4+960a^2\nu^2)\mu_8^5\\
 &\!\!\!\!+(234\nu^5+2816a^2\nu^3-44160a^4\nu)\mu_8^4+(116\nu^6+4032a^2\nu^4-256384a^4\nu^2)\mu_8^3\\
 &\!\!\!\!\!\!\!+(24\nu^7+2816a^2\nu^5-567552a^4\nu^3-3548160a^6\nu)\mu_8^2+(768a^2\nu^6-570368a^4\nu^4\\
 &\!\!-10407936a^6\nu^2)\mu_8+(-215040a^4\nu^5-7454720a^6\nu^3-92897280a^8\nu)\;.
\end{eqnarray*}
The respective roots with largest positive real part behave
asymptotically as:
\[\mu_4\sim\left\{\begin{array}{ll}
 \displaystyle24\frac{a^2}{\nu}-576\frac{a^4}{\nu^3}&\displaystyle\textrm{for }\frac{a}{\nu}\to0\\[0.3cm]
 \displaystyle2\sqrt[5]{36a^4\nu}&\displaystyle\textrm{for }\frac{a}{\nu}\to\infty\;,
\end{array}\right.\]
\[\mu_6\sim\left\{\begin{array}{ll}
 \displaystyle48\frac{a^2}{\nu}-2016\frac{a^4}{\nu^3}&\displaystyle\textrm{for }\frac{a}{\nu}\to0\\[0.3cm]
 \displaystyle2\sqrt[7]{1800a^6\nu}&\displaystyle\textrm{for }\frac{a}{\nu}\to\infty\;,
\end{array}\right.\]
\[\mu_8\sim\left\{\begin{array}{ll}
 \displaystyle80\frac{a^2}{\nu}-5280\frac{a^4}{\nu^3}&\displaystyle\textrm{for }\frac{a}{\nu}\to0\\[0.3cm]
 \displaystyle2\sqrt[9]{181440a^8\nu}&\displaystyle\textrm{for }\frac{a}{\nu}\to\infty\;.
\end{array}\right.\]
These relations provide us with three points of the curve
$\gamma(n)$ in the asymptotic conditions, to which three more can
be added: besides the already-mentioned $\gamma(2)$, one should
indeed remember that both $\gamma(0)$ and $\gamma(-2)$ (in 2D)
must vanish \cite{F63,ZRMS84,FGV01}. A simple extrapolation then
suggests the following asymptotic behaviours:
\begin{equation} \label{ggg}
 \gamma(n)\sim\left\{\begin{array}{ll}
  \displaystyle G(n)\frac{a^2}{\nu}+\mathcal{G}(n)\frac{a^4}{\nu^3}&\displaystyle\textrm{for }\frac{a}{\nu}\to0\\[0.3cm]
  \displaystyle g(n)a^{n/(n+1)}\nu^{1/(n+1)}&\displaystyle\textrm{for }\frac{a}{\nu}\to\infty\;.
 \end{array}\right.
\end{equation}
The coefficient $G(n)$ can be proven rigorously to be $n(n+2)$
(see the next subsection), while such a proof does not look
feasible for $\mathcal{G}(n)$. However, starting from the
quadratic form of the former, one can guess a fourth-order
polynomial for the latter, $\mathcal{G}(n)=3n(n+2)(n^2+2n+8)/4$,
and the correctness of this expression seems to be confirmed by
the fact that it satisfies the six conditions in $n=-2,0,2,4,6,8$
despite possessing only five degrees of freedom. As the following
coefficients in the power expansion at small $a/\nu$ are expected
to show higher powers of $n$, such an expansion is not uniform in
the sense that it must fail for some (large enough) value of $n$.
On the contrary, the coefficient $g(n)$ must originate from some
combinatorics, but has not been identified yet, while the presence
of non-integer powers of $a$ and $\nu$ suggests that the
large-$a/\nu$ development is singular. It is worth noticing the
dependence of the growth rate of the $n$-th moment as a function
of the noise correlation time (inverse of $\nu$): we have a linear
growth when the latter is small and a power-law decay when it is
large.\\
From (\ref{ggg}), exploiting the knowledge of $G(n)$ and the
vanishing of $g(0)$, we get the asymptotic behaviours of the
(positive, as expected) Lyapunov exponent:
\begin{equation} \label{lam}
 \lambda=\left\{\begin{array}{ll}
  \displaystyle2\frac{a^2}{\nu}&\displaystyle\textrm{for }\frac{a}{\nu}\to0\\[0.3cm]
  \displaystyle g'(0)\nu&\displaystyle\textrm{for }\frac{a}{\nu}\to\infty\;.
 \end{array}\right.
\end{equation}

\subsubsection{Quasi delta-correlated case} \label{qdcc}
The results obtained in the previous subsection are exact, but
they have been obtained for specific values of $n$, and the
generalization of the coefficients (\ref{ggg}) as functions of $n$
was empirical. There is however at least one instance in which
rigorous analytical results can be obtained: it is provided by the
quasi delta-correlated case, corresponding to small $a/\nu$.\\
It is worth reminding that in the exactly delta-correlated case
\cite{FGV01,BF99} the probability distribution of $R(t)$ is
lognormal, the Cram\'er function is quadratic and the Lyapunov
exponent is given by $2D_1$, where $D_1$ is the multiplicative
constant in the well-known expression for the spatial part of the
two-point structure function of the velocity field. A comparison
with (\ref{lam}) suggests the correspondence
$a^2/\nu\leftrightarrow D_1$, which means that the correct limit
is $a\to\infty\gets\nu$ such that $a/\nu\to0$ but $a^2/\nu$ keeps
finite.\\
The key point in this limit is that the leading behaviour can be
extracted rigorously, for any natural $n$, from a restricted
dynamics, namely keeping into account only the first $4(n+1)$
components in $\mathcal{R}^{(n)}$ (corresponding to the average of
the coordinates $r_{n,k}$, and of the latter times one noise). To
show this, for any quantity $Y$ let us define its small-Kubo
expansion as
\begin{equation} \label{series}
 Y=\sum_{l=0}^{\infty}\left(\frac{a}{\nu}\right)^lY_{[l]}
\end{equation}
and let us plug it in the evolution equation (\ref{ra}).\\
At the lowest order ($l=0$), we obtain
\begin{equation} \label{eqdiff}
 \ud_t\langle\mathcal{R}^{(n)}_{[0]\iota}\rangle=\left\{\begin{array}{ll}
  0&\textrm{for }1\le\iota\le n+1\\
  -\nu\langle\mathcal{R}^{(n)}_{[0]\iota}\rangle&\textrm{for }(n+1)+1\le\iota\le4(n+1)\\
  -2\nu\langle\mathcal{R}^{(n)}_{[0]\iota}\rangle&\textrm{for }4(n+1)+1\le\iota\le7(n+1)\\
  -3\nu\langle\mathcal{R}^{(n)}_{[0]\iota}\rangle&\textrm{for }7(n+1)+1\le\iota\le8(n+1)\;,
 \end{array}\right.
\end{equation}
so that $\langle\mathcal{R}^{(n)}_{[0]}\rangle$ is a constant in
the group of its first $n+1$ components (corresponding to the
average of the coordinates $r_{n,k}$) and decreases exponentially
in the remaining three groups.\\
The equations for $l=1$ have the same operatorial structure as in
(\ref{eqdiff}) but, instead of being homogeneous, are rather
forced by the aforementioned $l=0$ quantities in a ``cross
fashion'', i.e.~not by quantities of the same group but of
``nearest-neighbour'' groups. Therefore, the only nondecreasing
source term ($\langle\mathcal{R}^{(n)}_{[0]\iota}\rangle$ with
$1\le\iota\le n+1$) appears exclusively in the evolution equations
for $\langle\mathcal{R}^{(n)}_{[1]\kappa}\rangle$ with
$(n+1)+1\le\kappa\le4(n+1)$, which implies that now
$\langle\mathcal{R}^{(n)}_{[1]}\rangle$ evolves towards a finite
constant in its first $4(n+1)$ components (corresponding to the
average of the coordinates $r_{n,k}$, and of the latter times one
noise) and decreases exponentially in the remaining ones.\\
Moving to $l=2$, the first $7(n+1)$ components are now forced by a
non-vanishing (``constant'') source term. This means that
$\langle\mathcal{R}^{(n)}_{[2]\iota}\rangle$ goes to a constant
for $(n+1)+1\le\iota\le7(n+1)$, while the first $n+1$ components
(corresponding to the average of the coordinates $r_{n,k}$) get a
linear behaviour in time. For $l=3$ also the components with
$(n+1)+1\le\iota\le4(n+1)$ behave linearly, which in turn implies
a quadratic behaviour for the first $n+1$ components at $l=4$, and
so on.\\
The situation is resumed in the following table, in which for
every $l$ we show the degree $T$ of the leading behaviour in time
of each group of components
$\langle\mathcal{R}^{(n)}_{[2]\iota}\rangle$. E.g., 0 denotes the
evolution towards a finite constant, 1 a linear behaviour and
$\times$ the only presence of a decreasing exponential.\\[0.5cm]
\begin{tabular}{p{4cm}|cccccccc}
 $\hspace{3.5cm}l=$&0&1&2&3&4&5&6\\
 \hline\hline
 $1\le\iota\le n+1\hspace{1cm}$ i.e.~$\langle r_{n,k}\rangle$&0&0&1&1&2&2&3&\ldots\\
 \hline
 $(n+1)+1\le\iota\le4(n+1)$ i.e.~$\langle\xi_{\bullet}r_{n,k}\rangle$&$\times$&0&0&1&1&2&2&\ldots\\
 \hline
 $4(n+1)+1\le\iota\le7(n+1)$ i.e.~$\langle\xi_{\bullet}\xi_{\circ} r_{n,k}\rangle$&$\times$&$\times$&0&0&1&1&2&\ldots\\
 \hline
 $7(n+1)+1\le\iota\le8(n+1)$ i.e.~$\langle\altsu\altsd\altst r_{n,k}\rangle$&$\times$&$\times$&$\times$&0&0&1&1&\ldots
\end{tabular}\\[0.5cm]
Analysing the first line of this table, we argue that the expected
exponential behaviour for $\langle r_{n,k}\rangle$ corresponds to
the summation of the series in (\ref{series}) in the fashion
\begin{equation} \label{ser}
 \langle r_{n,k}(t)\rangle=\sum_{L=0}^{\infty}\left(\frac{a}{\nu}\right)^{2L}(\nu t)^Lf(n,k,L)\left[1+O\left(\frac{a}{\nu}\right)\right]\stackrel{t\to+\infty}{\sim}\ue^{\mu_nt}\textrm{ (for some }k)
\end{equation}
One would be tempted to conclude that, for each $n$, the
largest-positive-real-part eigenvalue $\mu_n$ could be easily
extracted by looking at the coefficient in (\ref{ser}) with $L=1$,
corresponding to the column $l=2$ in the table, because
$\ue^{c\varepsilon}\sim1+c\varepsilon$ for small $\varepsilon$.
However, this would be the case only if the eigenvalues
corresponding to a fixed $n$ were definitely separate for
$a/\nu\to0$. On the contrary, from the study of the full matrix
performed e.g.~for $n=2$ (see also appendix \ref{app2}), we know
that several eigenvalues vanish in this limit, and we have to
identify the one which keeps the largest positive real part in
this process: we are therefore facing a \emph{degenerate
perturbation theory}, and the quest for the asymptotic expression
of $\mu_n$ (or, in a way, for the rigorous proof that it has the
expression shown in the previous subsection) is somehow longer.\\
Let us then proceed as follows. From the inspection of the system
(\ref{eqdiff}) and of the following orders, it is apparent that
the structure of the equations is such that the degree $T$ of the
leading behaviour in time (i.e.~the quantity reported in the
table) can grow of one unity only when a quantity of the second
group, of order $l$ and time degree $T$, acts as a source term for
a quantity of the first group of order $l+1$, which therefore gets
a degree $T+1$. This means that, in order to find the leading
behaviour of $\mu_n$ at small $a/\nu$, one can completely neglect
the lower two rows in the table and only consider the upper two.
Indeed, starting from the only non-vanishing group ($T=0$) in the
column $l=0$ and remembering that the forcing mechanism acts in a
diagonal ``cross fashion'' (i.e.~every term can only force its
upper-right and lower-right neighbours), the components of the
third and of the fourth group can give rise to linear terms in the
first group only for $l=4$ and $l=6$ respectively, which are
clearly subdominant with respect both to the quadratic and cubic
behaviours of these same components ($T=2$ and $T=3$ in the table)
and to the linear behaviour found for $l=2$ due to the forcing by
the second group. Back to the initial dynamics
$\ud_t\langle\mathcal{R}^{(n)}\rangle=\ldots$, this amounts to say
that in the evolution equation for $\langle r_{n,k}\rangle$ one
should keep the complete form into account, while for
$\langle\xi_{\bullet}r_{n,k}\rangle$ one can neglect all source
terms involving two different noises: the system thus closes at
this stage, reducing the order of the corresponding matrix from
the initial $8(n+1)$ to $4(n+1)$. In other words, only the
upper-left quarter of the matrix $\mathtt{A}^{(n)}$ is relevant.\\
The problem can now easily be recast as a system of $n+1$
second-order differential equations for $r_{n,k}$. Indeed, we have
\begin{eqnarray*}
 \ud^2_tr_{n,k}\!\!&\!\!=\!\!&\!\!kR_1^{k-1}R_2^{n-k}\ud^2_tR_1+(n-k)R_1^kR_2^{n-k-1}\ud^2_tR_2\\
 &&\!\!+k(k-1)R_1^{k-2}R_2^{n-k}(\ud_tR_1)^2+(n-k)(n-k-1)R_1^kR_2^{n-k-2}(\ud_tR_2)^2\\
 &&\!\!+2k(n-k)R_1^{k-1}R_2^{n-k-1}\ud_tR_1\ud_tR_2\;,
\end{eqnarray*}
and, keeping into account the aforementioned simplification in the
equations for $\langle\xi_{\bullet}r_{n,k}\rangle$,
\begin{eqnarray} \label{secord}
 \ud^2_t\langle r_{n,k}\rangle+\nu\ud_t\langle r_{n,k}\rangle\!\!&\!\!=\!\!&\!\!a^2\left[(n^2+6k^2-6nk-n)\langle r_{n,k}\rangle\right.\\
 &&\!\!+3k(k-1)\langle r_{n,k-2}\rangle+3(n-k)(n-k-1)\langle r_{n,k+2}\rangle\big]\nonumber\\
 &\!\!\equiv\!\!&\!\!a^2\mathsf{M}^{(n)}_{kk'}\langle r_{n,k'}\rangle\;.\nonumber
\end{eqnarray}
The matrix $\mathsf{M}^{(n)}$, defined by the right-hand side of
(\ref{secord}), is a priori of order $(n+1)\times(n+1)$. However,
it is apparent from (\ref{secord}) that, for a fixed $n$, the
dynamics of the components of $\langle r_{n,k}\rangle$ with even
$k$ are independent of those with odd $k$. Therefore, as our main
goal is to reconstruct objects like (\ref{binom}), we can just
focus on subset $\hat{\mathsf{M}}^{(n)}$ of $\mathsf{M}^{(n)}$
consisting of the lines with even $k$, and thus reduce to a
$(n/2+1)\times(n/2+1)$ problem, simply neglecting the lines
corresponding to odd $k$. In other words, we rewrite
(\ref{secord}) as
\begin{equation} \label{hat}
 \ud^2_t\langle r_{n,2k}\rangle+\nu\ud_t\langle r_{n,2k}\rangle=a^2\sum_{k'=0,2,\ldots,n}\mathsf{M}^{(n)}_{2k,k'}\langle r_{n,k'}\rangle=a^2\hat{\mathsf{M}}^{(n)}_{kk'}\langle r_{n,2k'}\rangle\;.
\end{equation}
Moreover, interpreting (\ref{hat}) as
\[\left\{\begin{array}{l}
 \ud_t\langle r_{n,2k}\rangle=-\nu\langle r_{n,2k}\rangle+\nu Y_{n,2k}\\
 \displaystyle\ud_tY_{n,2k}=\frac{a^2}{\nu}\hat{\mathsf{M}}^{(n)}_{kk'}\langle r_{n,2k'}\rangle\;,
\end{array}\right.\]
where $Y_{n,2k}\equiv\langle r_{n,2k}\rangle+\ud_t\langle
r_{n,2k}\rangle/\nu$, it is easy to show that the
largest-positive-real-part eigenvalue $\mu_n$ (divided by
$a^2/\nu$) that we are looking for must belong to the spectrum of
the subset matrix $\hat{\mathsf{M}}^{(n)}$, because the remaining
eigenvalues are all $-\nu$. Such subset matrix is tridiagonal
centrosymmetric but not symmetric, which means that left and right
eigenvectors differ. It is a simple task to prove that the value
$n(n+2)$ (guessed for $\mu_2/(a^2/\nu)$ asymptotically from the
previous subsection) is actually an eigenvalue, but it is not
possible to use a variational method to show that it has the
largest positive real part. However, this is not needed, because
the row vector built with the $n/2+1$ components of the binomial
coefficient $\binom{n/2}{k}$, $v_k\equiv(1,n/2,\ldots,n/2,1)$ for
$k=0,1,\ldots,n/2$, happens to be the left eigenvector of
$\hat{\mathsf{M}}^{(n)}$ corresponding to the eigenvalue $n(n+2)$.
Therefore, using (\ref{binom}) and (\ref{hat}), one has:
\begin{eqnarray*}
 \ud^2_t\langle R^n\rangle+\nu\ud_t\langle R^n\rangle\!\!&\!\!=\!\!&\!\!\sum_{k=0,2,\ldots,n}\binom{n/2}{k/2}\left(\ud^2_t\langle r_{n,k}\rangle+\nu\ud_t\langle r_{n,k}\rangle\right)\\
 &\!\!=\!\!&\!\!\sum_{k=0,1,\ldots,n/2}\binom{n/2}{k}\left(\ud^2_t\langle r_{n,2k}\rangle+\nu\ud_t\langle r_{n,2k}\rangle\right)\\
 &\!\!=\!\!&\!\!\sum_{k=0,1,\ldots,n/2}v_ka^2\sum_{k'=0,1,\ldots,n/2}\hat{\mathsf{M}}^{(n)}_{kk'}\langle r_{n,2k'}\rangle\\
 &\!\!=\!\!&\!\!a^2\sum_{k'=0,1,\ldots,n/2}\left(\sum_{k=0,1,\ldots,n/2}v_k\hat{\mathsf{M}}^{(n)}_{kk'}\right)\langle r_{n,2k'}\rangle\\
 &\!\!=\!\!&\!\!a^2\sum_{k'=0,1,\ldots,n/2}n(n+2)v_{k'}\langle r_{n,2k'}\rangle\\
 &\!\!=\!\!&\!\!n(n+2)a^2\sum_{k'=0,1,\ldots,n/2}\binom{n/2}{k'}\langle r_{n,2k'}\rangle\\
 &\!\!=\!\!&\!\!n(n+2)a^2\sum_{k'=0,2,\ldots,n}\binom{n/2}{k'/2}\langle r_{n,k'}\rangle\\
 &\!\!=\!\!&\!\!n(n+2)a^2\langle R^n\rangle\;.
\end{eqnarray*}
This implies
\[\mu_n\stackrel{a/\nu\to0}{\sim}\frac{\sqrt{\nu^2+4n(n+2)a^2}-\nu}{2}=n(n+2)\frac{a^2}{\nu}\left[1+O\left(\frac{a}{\nu}\right)^2\right]\]
as expected.\\
Notice that the next-leading correction cannot be captured at this
stage, because of the approximation introduced previously. To take
it into account correctly, one should reformulate the analysis
performed in this section, including also quantities like
$\langle\xi_{\bullet}\xi_{\circ}r_{n,k}\rangle$ but however
excluding $\langle\altsu\altsd\altst r_{n,k}\rangle$,
i.e.~considering a reduced dynamics in the first $7(n+1)$
components of
$\mathcal{R}^{(n)}$.\\
It is worth mentioning that an analysis like the previous one, but
applied to odd $n$'s, leads to
\[\mu_n\stackrel{a/\nu\to0}{\sim}(n-1)(n+3)\frac{a^2}{\nu}\;,\]
in accordance e.g.~with the exact result $\mu_1=0$ found with the
complete dynamics. However, as already pointed out, such values
are not related to the curve $\gamma(n)$.

Lastly, one would be interested in reformulating the previous
study in the quasi-deterministic limit, $a/\nu\to\infty$ (see
appendix \ref{app1}). Unfortunately, this is not possible, because
the limit turns out to be singular: an exponential growth is
expected in general for finite $a/\nu$, but a power law is found
when this ratio is infinite. In other words, an expansion like
(\ref{series}) with $a/\nu$ inverted does not work. In (\ref{ggg})
this reflects into the presence of non-integer powers and into the
difference between the limits $a\to\infty$ and $\nu\to0$.

\section{Conclusions} \label{conc}

By assuming the telegraph-noise model for the velocity gradient
(or strain matrix), we were able to carry out analytical
computations and to obtain several results on the separation
between two fluid particles. Focusing on smooth flows (Batchelor
regime), we firstly analysed the one-dimensional compressible
case, finding explicit expressions for the long-time evolution of
the interparticle-distance moments (\ref{alfagamma}), the Lyapunov
exponent (\ref{lambda}) and the Cram\'er function (\ref{cramer}).
Then we concentrated on the two-dimensional incompressible
situation and provided an elementary extension of the 1D case,
represented by the hyperbolic flow. Moving to the general
isotropic case, a thorough analysis on the complete dynamics was
made for the evolution of linear ($n=1$) and quadratic ($n=2$)
components. Due to high computational cost, at higher $n$ we
focused on a restricted, though exact, dynamics: however, only
specific values of $n$ could be studied in this way, leading to
the extrapolations (\ref{ggg}) and (\ref{lam}). Such guess was
rigorously proved in the quasi-delta-correlated limit, for which
approximated equations were introduced.

We believe that the present paper represents an interesting
example of the use of a coloured noise for which the well-known
closure problem can be solved analytically \cite{FMA07}.

\vfill

\section*{Acknowledgements}

I warmly thank Grisha Falkovich, who inspired this work. I
acknowledge useful discussions with Itzhak Fouxon, Vladimir
Lebedev, Stefano Musacchio, Konstantin Turitsyn and Marija
Vucelja.

\vfill

\appendix

\section{Appendix: deterministic case in 2D} \label{app1}

The deterministic case corresponds to $\nu=0$, when the noise
$\xi(t)$ takes a constant value $\pm a$. In 2D, the presence or
absence of rotation plays a crucial role. Indeed, in the diagonal
case (corresponding to the hyperbolic flow shown in section
\ref{hypf}), one finds the exponential evolution
$R^n\stackrel{t\to\infty}{\sim}\ue^{ant}$, which implies
$\gamma(n)=an$ and $\lambda=a$.\\
On the contrary, in the general isotropic case (with rotation),
one gets $\ud^2_tR_1=0=\ud^2_tR_2$ (for any of the $2^3=8$
possible combinations of the noise signs), thus both components
are linear in time and $R^n\stackrel{t\to\infty}{\sim}t^n$.
Consequently, a power-law temporal dependence is found for the
separation and $\lambda=0$. It is worth noticing that this result
is due to the fact that, with our choice, $\sigma^2=0$ and is
characteristic of the 2D isotropic situation.

\section{Appendix: calculation details for 2D} \label{app2}

In this appendix we provide some details of the calculation for
the 2D general isotropic case.\\
The matrix $\mathtt{A}^{(n)}_{\iota\kappa}$ reads:
\[\left\{\begin{array}{lr}
[(n+1)-2\iota+1]a\delta_{\kappa,\iota+(n+1)}+[(n+1)-\iota]a\delta_{\kappa,\iota+2(n+1)+1}&\\
 +\sqrt{2}[(n+1)-\iota]a\delta_{\kappa,\iota+3(n+1)+1}+(\iota-1)a\delta_{\kappa,\iota+2(n+1)-1}&\\
 -\sqrt{2}(\iota-1)a\delta_{\kappa,\iota+3(n+1)-1}&\\
 &\hspace{-5cm}\textrm{for }1\le\iota\le n+1\;,\\[0.2cm]
[3(n+1)-2\iota+1]a\delta_{\kappa,\iota-(n+1)}+[2(n+1)-\iota]a\delta_{\kappa,\iota+3(n+1)+1}&\\
 +\sqrt{2}[2(n+1)-\iota]a\delta_{\kappa,\iota+4(n+1)+1}+[-(n+1)+\iota-1]a\delta_{\kappa,\iota+3(n+1)-1}&\\
 -\sqrt{2}[-(n+1)+\iota-1]a\delta_{\kappa,\iota+4(n+1)-1}-\nu\delta_{\kappa,\iota}&\\
 &\hspace{-5cm}\textrm{for }(n+1)+1\le\iota\le2(n+1)\;,\\[0.2cm]
[5(n+1)-2\iota+1]a\delta_{\kappa,\iota+2(n+1)}+[3(n+1)-\iota]a\delta_{\kappa,\iota-2(n+1)+1}&\\
 +\sqrt{2}[3(n+1)-\iota]a\delta_{\kappa,\iota+4(n+1)+1}+[-2(n+1)+\iota-1]a\delta_{\kappa,\iota-2(n+1)-1}&\\
 -\sqrt{2}[-2(n+1)+\iota-1]a\delta_{\kappa,\iota+4(n+1)-1}-\nu\delta_{\kappa,\iota}&\\
 &\hspace{-5cm}\textrm{for }2(n+1)+1\le\iota\le3(n+1)\;,\\[0.2cm]
[7(n+1)-2\iota+1]a\delta_{\kappa,\iota+2(n+1)}+[4(n+1)-\iota]a\delta_{\kappa,\iota+3(n+1)+1}&\\
 +\sqrt{2}[4(n+1)-\iota]a\delta_{\kappa,\iota-3(n+1)+1}+[-3(n+1)+\iota-1]a\delta_{\kappa,\iota+3(n+1)-1}&\\
 -\sqrt{2}[-3(n+1)+\iota-1]a\delta_{\kappa,\iota-3(n+1)-1}-\nu\delta_{\kappa,\iota}&\\
 &\hspace{-5cm}\textrm{for }3(n+1)+1\le\iota\le4(n+1)\;,\\[0.2cm]
[9(n+1)-2\iota+1]a\delta_{\kappa,\iota-2(n+1)}+[5(n+1)-\iota]a\delta_{\kappa,\iota-3(n+1)+1}&\\
 +\sqrt{2}[5(n+1)-\iota]a\delta_{\kappa,\iota+3(n+1)+1}+[-4(n+1)+\iota-1]a\delta_{\kappa,\iota-3(n+1)-1}&\\
 -\sqrt{2}[-4(n+1)+\iota-1]a\delta_{\kappa,\iota+3(n+1)-1}-2\nu\delta_{\kappa,\iota}&\\
 &\hspace{-5cm}\textrm{for }4(n+1)+1\le\iota\le5(n+1)\;,\\[0.2cm]
[11(n+1)-2\iota+1]a\delta_{\kappa,\iota-2(n+1)}+[6(n+1)-\iota]a\delta_{\kappa,\iota+2(n+1)+1}&\\
 +\sqrt{2}[6(n+1)-\iota]a\delta_{\kappa,\iota-4(n+1)+1}+[-5(n+1)+\iota-1]a\delta_{\kappa,\iota+2(n+1)-1}&\\
 -\sqrt{2}[-5(n+1)+\iota-1]a\delta_{\kappa,\iota-4(n+1)-1}-2\nu\delta_{\kappa,\iota}&\\
 &\hspace{-5cm}\textrm{for }5(n+1)+1\le\iota\le6(n+1)\;,\\[0.2cm]
[13(n+1)-2\iota+1]a\delta_{\kappa,\iota+(n+1)}+[7(n+1)-\iota]a\delta_{\kappa,\iota-3(n+1)+1}&\\
 +\sqrt{2}[7(n+1)-\iota]a\delta_{\kappa,\iota-4(n+1)+1}+[-6(n+1)+\iota-1]a\delta_{\kappa,\iota-3(n+1)-1}&\\
 -\sqrt{2}[-6(n+1)+\iota-1]a\delta_{\kappa,\iota-4(n+1)-1}-2\nu\delta_{\kappa,\iota}&\\
 &\hspace{-5cm}\textrm{for }6(n+1)+1\le\iota\le7(n+1)\;,\\[0.2cm]
[15(n+1)-2\iota+1]a\delta_{\kappa,\iota-(n+1)}+[8(n+1)-\iota]a\delta_{\kappa,\iota-2(n+1)+1}&\\
 +\sqrt{2}[8(n+1)-\iota]a\delta_{\kappa,\iota-3(n+1)+1}+[-7(n+1)+\iota-1]a\delta_{\kappa,\iota-2(n+1)-1}&\\
 -\sqrt{2}[-7(n+1)+\iota-1]a\delta_{\kappa,\iota-3(n+1)-1}-2\nu\delta_{\kappa,\iota}&\\
 &\hspace{-5cm}\textrm{for }7(n+1)+1\le\iota\le8(n+1)\;.
\end{array}\right.\]

We focus at first on the study of linear coordinates ($n=1$): we
introduce the 16-component vector
\[\mathcal{R}^{(1)}\equiv\left(\begin{array}{c}R_1\\R_2\\\altsu(t)R_1/a\\\altsu(t)R_2/a\\\altsd(t)R_1/a\\\altsd(t)R_2/a\\\altst(t)R_1/a\\\altst(t)R_2/a\\\altsu(t)\altsd(t)R_1/a^2\\\altsu(t)\altsd(t)R_2/a^2\\\altsu(t)\altst(t)R_1/a^2\\\altsu(t)\altst(t)R_2/a^2\\\altsd(t)\altst(t)R_1/a^2\\\altsd(t)\altst(t)R_2/a^2\\\altsu(t)\altsd(t)\altst(t)R_1/a^3\\\altsu(t)\altsd(t)\altst(t)R_2/a^3\end{array}\right)\;.\]
The matrix $\mathtt{A}^{(1)}$ reads
\[{\setlength\arraycolsep{0pt}\left(\begin{array}{cccccccccccccccc}
 0&0&a&0&0&a&0&\sqrt{2}a&0&0&0&0&0&0&0&0\\
 0&0&0&-a&a&0&-\sqrt{2}a&0&0&0&0&0&0&0&0&0\\
 a&0&-\nu&0&0&0&0&0&0&a&0&\sqrt{2}a&0&0&0&0\\
 0&-a&0&-\nu&0&0&0&0&a&0&-\sqrt{2}a&0&0&0&0&0\\
 0&a&0&0&-\nu&0&0&0&a&0&0&0&0&\sqrt{2}a&0&0\\
 a&0&0&0&0&-\nu&0&0&0&-a&0&0&-\sqrt{2}a&0&0&0\\
 0&\sqrt{2}a&0&0&0&0&-\nu&0&0&0&a&0&0&a&0&0\\
 -\sqrt{2}a&0&0&0&0&0&0&-\nu&0&0&0&-a&a&0&0&0\\
 0&0&0&a&a&0&0&0&-2\nu&0&0&0&0&0&0&\sqrt{2}a\\
 0&0&a&0&0&-a&0&0&0&-2\nu&0&0&0&0&-\sqrt{2}a&0\\
 0&0&0&\sqrt{2}a&0&0&a&0&0&0&-2\nu&0&0&0&0&a\\
 0&0&-\sqrt{2}a&0&0&0&0&-a&0&0&0&-2\nu&0&0&a&0\\
 0&0&0&0&0&\sqrt{2}a&0&a&0&0&0&0&-2\nu&0&a&0\\
 0&0&0&0&-\sqrt{2}a&0&a&0&0&0&0&0&0&-2\nu&0&-a\\
 0&0&0&0&0&0&0&0&0&\sqrt{2}a&0&a&a&0&-3\nu&0\\
 0&0&0&0&0&0&0&0&-\sqrt{2}a&0&a&0&0&-a&0&-3\nu
\end{array}\right)}\]
and its eigenvalues are given by $0$ (twice), $-\nu$ (six times),
$-2\nu$ (six times) and $-3\nu$ (twice). Therefore, the largest
eigenvalue is $\mu_1=0$, with eigenvectors
$\left(\begin{array}{c}\nu/a\\0\\1\\0\\0\\1\\0\\-\sqrt{2}\\0\\0\\0\\0\\0\\0\\0\\0\end{array}\right)$
and
$\left(\begin{array}{c}0\\\nu/a\\0\\-1\\1\\0\\\sqrt{2}\\0\\0\\0\\0\\0\\0\\0\\0\\0\end{array}\right)$.\\[0.5cm]

Let us now move to quadratic quantities ($n=2$): we define the
24-component vector
\[\mathcal{R}^{(2)}\equiv\left(\begin{array}{c}R_1^2\\R_1R_2\\R_2^2\\\altsu(t)R_1^2/a\\\altsu(t)R_1R_2/a\\\altsu(t)R_2^2/a\\\altsd(t)R_1^2/a\\\altsd(t)R_1R_2/a\\\altsd(t)R_2^2/a\\\altst(t)R_1^2/a\\\altst(t)R_1R_2/a\\\altst(t)R_2^2/a\\\altsu(t)\altsd(t)R_1^2/a^2\\\altsu(t)\altsd(t)R_1R_2/a^2\\\altsu(t)\altsd(t)R_2^2/a^2\\\altsu(t)\altst(t)R_1^2/a^2\\\altsu(t)\altst(t)R_1R_2/a^2\\\altsu(t)\altst(t)R_2^2/a^2\\\altsd(t)\altst(t)R_1^2/a^2\\\altsd(t)\altst(t)R_1R_2/a^2\\\altsd(t)\altst(t)R_2^2/a^2\\\altsu(t)\altsd(t)\altst(t)R_1^2/a^3\\\altsu(t)\altsd(t)\altst(t)R_1R_2/a^3\\\altsu(t)\altsd(t)\altst(t)R_2^2/a^3\end{array}\right)\;.\]
The associated matrix is
\newpage\thispagestyle{empty}
\begin{figure}[h!]
 \begin{picture}(300,400)
  \put(0,450){\includegraphics[angle=180]{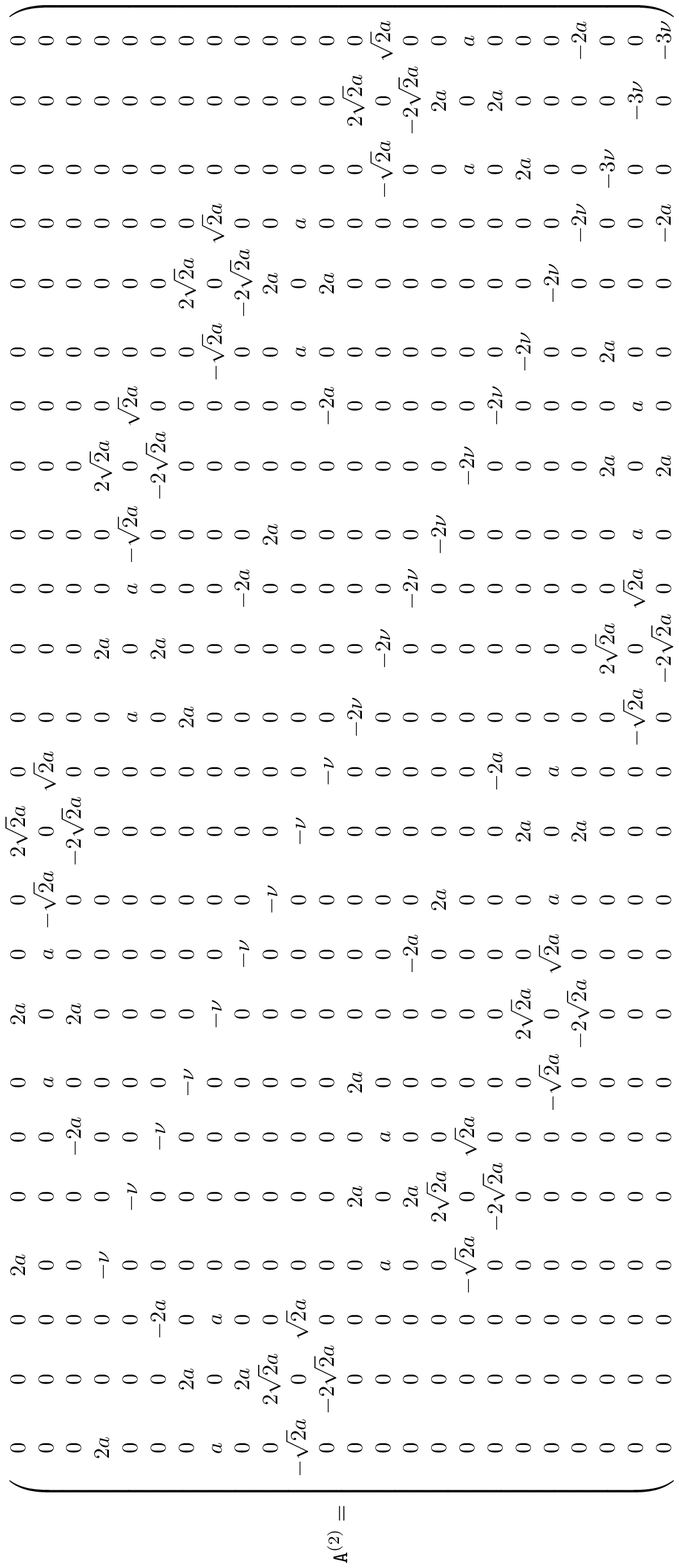}}
 \end{picture}
\end{figure}\newpage
\noindent and its eigenvalues are: $-2\nu$ (three times), $-\nu$
(three times), the three roots of
$\mu_2^3+3\nu\mu_2^2+2\nu^2\mu_2-16a^2\nu=0$ (corresponding to the
reduced dynamics (\ref{red})), the three roots (each taken twice)
of $\mu_2^3+3\nu\mu_2^2+2\nu^2\mu_2+8a^2\nu=0$, the three roots
(each taken twice) of
$\mu_2^3+6\nu\mu_2^2+11\nu^2\mu_2+(6\nu^3-8a^2\nu)=0$ and the
three roots of
$\mu_2^3+6\nu\mu_2^2+11\nu^2\mu_2+(6\nu^3+16a^2\nu)=0$. In
accordance with (\ref{mu}), the largest eigenvalue is
\[\mu_2=-\nu+\frac{{\nu}^2}{\sqrt[3]{216a^2\nu+3\sqrt{5184a^4\nu^2-3\nu^6}}}+\frac{\sqrt[3]{216a^2\nu+3\sqrt{5184a^4\nu^2-3\nu^6}}}{3}\]
and its eigenvector reads:
\[\left(\begin{array}{c}\sqrt{2}+2\sqrt{2}\nu/\mu\\0\\\sqrt{2}+2\sqrt{2}\nu/\mu\\\nu/\sqrt{2}a+\mu/2\sqrt{2}a\\0\\-\nu/\sqrt{2}a-\mu/2\sqrt{2}a\\0\\\nu/\sqrt{2}a+\mu/2\sqrt{2}a\\0\\0\\0\\0\\0\\0\\0\\0\\-1\\0\\1\\0\\-1\\0\\0\\0\end{array}\right)\;.\]

\end{document}